\documentclass[aps,twocolumn,showpacs]{revtex4}
\newcommand{\sect}[1]{\setcounter{equation}{0}\section{#1}}

\begin{document}
\title{Statistical mechanics in the context of special relativity II}
\author{G. Kaniadakis}
\email{giorgio.kaniadakis@polito.it} \affiliation{Dipartimento di
Fisica, Politecnico di Torino, \\ Corso Duca degli Abruzzi 24,
10129 Torino, Italy}
\date{\today}

\begin{abstract}
The special relativity laws emerge as one-parameter (light speed)
generalizations of the corresponding laws of classical physics.
These generalizations, imposed by the Lorentz transformations,
affect both the definition of the various physical observables
(e.g. momentum, energy, etc), as well as the mathematical
apparatus of the theory. Here, following the general lines of
[Phys. Rev. E {\bf 66}, 056125 (2002)], we show that the Lorentz
transformations impose also a proper one-parameter
generalization  of the classical Boltzmann-Gibbs-Shannon entropy.
The obtained relativistic entropy permits to construct a coherent
and selfconsistent relativistic statistical theory, preserving the
main features of the ordinary statistical theory, which recovers
in the classical limit. The predicted distribution function is a
one-parameter continuous deformation of the classical
Maxwell-Boltzmann distribution and has a simple analytic form,
showing power law tails in accordance with the experimental
evidence. Furthermore the new statistical mechanics can be
obtained as stationary case of a generalized kinetic theory
governed by an evolution equation obeying the H-theorem and
reproducing the Boltzmann equation of the ordinary kinetics in
the classical limit.

\end{abstract}
 \maketitle

\sect{Introduction}

In high energy physics, various approaches of statistical
mechanics and kinetic theory have been adopted to study
cosmological models \cite{Bernstein,Kremer}, multiparticle
production in hadron-hadron collisions \cite{Giovannini},
possible Lorentz invariance violation in extension of
standard-model \cite{Colladay}, neutron star matter \cite{Prix},
black hole models \cite{Hyun,Husain,Gour,Bhaduri} etc. Common to
all such approaches is the explicit or implicit assumption of a
very specific form of the entropy, i.e. the celebrated
Boltzmann-Gibbs-Shannon entropy which, according to Jaynes
\cite{Jaynes} maximum entropy, principle conducts to the
Maxwell-Boltzmann exponential distribution.

On the other hand since long time it is known that there exist
open problems in high energy physics which cannot be solved within
the ordinary Maxwell-Boltzmann statistics e.g  the cosmic rays
spectrum \cite{Vasyliunas,Biermann,Swordy} and the black hole
entropy Bekenstein-Hawking area law \cite{Bekenstein,Hawking}.

In statistical physics, beside the ubiquitous exponential
distribution $\exp (-x)$, also non-exponential distributions have
been considered. The most famous ones appear within the quantum
statistical mechanics, namely the Fermi-Dirac and the
Bose-Einstein distributions, which are forced by the
exclusion-inclusion principle. In the last decades a particular
attention has been addressed  to the statistical distributions
presenting power law tails as $x^{-a}$. The latter distributions
have been observed experimentally in many fields of physics,
including cosmic rays physics \cite{Vasyliunas,Biermann,Swordy},
plasma physics \cite{Hasegawa} and multiparticle production
processes \cite{Wilk}.

In order to propose theories based on non-exponential statistical
distributions, one can follow two different ways. The first is the
entropic one consisting of a proper generalization of the system
entropy. After fixing the entropy, the Jaynes maximum entropy
principle univocally determines the form of the distribution
function. The problem then is essentially reduced to the choice
of the right entropic form.

The second way is the kinetic one which is very familiar in non
equilibrium statistical mechanics. It is well known that the
classical statistical mechanics can be obtained as particular case
of the ordinary kinetic theory for $t\rightarrow\infty$, within
the Boltzmann or the Fokker-Planck picture. Generalized kinetic
theories necessarily conduct to generalized statistical mechanics
producing non-exponential distributions.

Both the entropic as well as the kinetic path, conduct to the
conclusion that it is possible to develop, beside the classical
statistical mechanics, other coherent and selfconsistent
statistical theories
\cite{PhA01,PRE02,PRE05LKS,Abe,Naudts,Chavanis,Bashkirov}.

It is remarkable that the Maxwell-Boltzmann distribution does not
emerge within the statistical mechanics. It is more proper to say
that the classical statistical mechanics is built  starting from
the Maxwell-Boltzmann distribution. It is commonly accepted that
this distribution emerges within the Newton mechanics. Indeed,
numerical simulation of classical molecular dynamics
unequivocally conducts to the exponential distribution. At this
point, the question if the Maxwell-Boltzmann distribution obtains
also in the case when the microscopic dynamics is governed by the
special relativity laws, naturally arises.

In the relativistic statistical theory the collision invariants
are imposed by the laws of special relativity. Hence the first
difference between the classical and  the relativistic
distribution functions regards the different forms of the
collision invariants, in the argument of the distribution
function. A second difference could be originated by the form of
the entropy of the relativistic many body system, which could be
different with respect to the Boltzmann-Gibbs-Shannon entropy.

In the ordinary relativistic statistical mechanics it is accepted
that the entropy has the same form  of the classical statistical
mechanics. Then the Jaynes principle yields an exponential
distribution which  reproduces exactly the Maxwell-Boltzmann
distribution in the rest frame. The most famous and important
relativistic particle system is undoubtedly the cosmic rays
\cite{Vasyliunas,Biermann,Swordy}. The cosmic rays spectrum has a
very large extension (13 decades in energy and 33 decades in
particle flux) and presents a power law asymptotic behavior.
Unfortunately the Maxwell-Boltzmann distribution fails to explain
this experimental spectrum. The same power law asymptotic
behavior of the distribution function was observed also in other
relativistic systems \cite{Hasegawa,Wilk}. Hence for relativistic
particle systems, the experimental evidence suggests a
non-exponential distribution function with power law tails. This
distribution, can be originated exclusively  by an entropy
manifestly different from the Boltzmann-Gibbs-Shannon one.

We recall that all the classical physical observables (momentum,
energy, etc) when considered within the special relativity are
properly generalized. Any relativistic formula can be viewed as a
one-parameter (light speed) generalization or deformation of the
corresponding classical formula. Consequently it is natural to
assume that also the entropy of a relativistic system could be a
one-parameter generalization of the classical entropy. In this
case the logarithm appearing in the expression of
Boltzmann-Gibbs-Shannon entropy must be replaced by another one
parameter generalized logarithm. Then also the ordinary
exponential function appearing in the Maxwell-Boltzmann
distribution must be replaced by an one-parameter generalized
exponential. These two generalized functions have been proposed
heuristically in ref. \cite{PhA01} and are given by
\begin{eqnarray}
&&\exp_{_{\{{\scriptstyle \kappa}\}}}(t)=\left(\sqrt{1+\kappa^2
t^2 }+ \kappa t \right)^{1/\kappa} \ , \label{RI1} \\
&&\ln_{_{\{{\scriptstyle \kappa}\}}} (t)=
\frac{t^{\kappa}-t^{-\kappa}}{2\kappa} \ . \label{RI2}
\end{eqnarray}
Note that in the classical limit $\kappa \rightarrow 0$ one easily
recovers the ordinary exponential and logarithm namely
$\exp_{_{\{{\scriptstyle 0}\}}}(t)=\exp (t)$, $ \ \ \
\ln_{_{\{{\scriptstyle 0}\}}}(t)=\ln (t)$.  Successively in ref.
\cite{PRE02} it has been shown that the statistical theory,
developed starting from these generalized functions, emerges
within the framework of the special relativity. The entropy for a
relativistic many body system assumes the form
\begin{eqnarray}
S(f\,)=-\int d^3p \,\,\,f\,\ln_{_{\{{\scriptstyle
\kappa}\}}}\!(f)= -\,\langle \ln_{_{\{{\scriptstyle
\kappa}\}}}\!(f) \rangle \ ,\label{RI3}
\end{eqnarray}
while the relevant statistical distribution in the rest frame is
given by
\begin{equation}
f=\alpha \exp_{_{\{{\scriptstyle \kappa}\}}}\!\left(-\,
\frac{W-\mu}{\lambda\,k_{_{B}}T}\right) \ \ , \label{RI4}
\end{equation}
$W$ being  the relativistic kinetic energy, while $\alpha$ and
$\lambda$ are two constants depending on $\kappa$. These two
latter relationships are one-parameter generalizations of the
Bolzmann-Gibbs-Shannon entropy and of the Maxwell-Boltzmann
distribution respectively, which are recovered in the classical
limit $\kappa \rightarrow 0$.

Main goal of the present contribution is to show that the
functions $\exp_{_{\{{\scriptstyle \kappa}\}}}\!(t)$ and
$\ln_{_{\{{\scriptstyle \kappa}\}}}\!(t)$,  leading to the
distribution function (\ref{RI4}) and to the entropy (\ref{RI3}),
emerge naturally and unequivocally within the special relativity
theory. These functions replace the ordinary exponential and
logarithm, relegating them at the native status of classical
functions. It is remarkable that the relativistic statistical
theory, based on the entropy (\ref{RI3}), predicts a power law
asymptotic behaviour to the spectrum of relativistic particles
according to the experimental evidence.

The paper is organized as follows:

In Sect. II we recall some concepts of relativistic dynamics
focusing our attention on the deformations in the mathematical
formalism enforced by the finite value of the light speed.

In Sect. III we show that the form of $\exp_{_{\{{\scriptstyle
\kappa}\}}}\!(t)$ can be obtained within the special relativity
by using elementary algebraic calculations, starting from the
dispersion relation of free relativistic particles.

In Sect. IV  we reobtain the exponential $\exp_{_{\{{\scriptstyle
\kappa}\}}}\!(t)$ starting from the concept of Lorentz invariant
integration.

In Sect. V starting from the three-dimension $\kappa$-sum enforced
by the Lorentz transformations we introduce the concept of
generalized differential.

In Sect. VI we derive the forms of $\ln_{_{\{{\scriptstyle
\kappa}\}}}\!(t)$ and $\exp_{_{\{{\scriptstyle \kappa}\}}}\!(t)$
within the special relativity, starting from the additivity law
of relativistic momenta.

In Sect. VII we recall some concepts related to maximum entropy
principle, in the context of a class of generalized statistical
theories.

In Sect. VIII we consider the explicit form of the distribution
function within the special relativity.

In Sect. IX we consider the main features of the statistical
mechanics and of the kinetic theory based on the entropy
(\ref{RI3}).

In Sect. X we consider some applications of the present theory to
specific physical problems and compare the results with the ones
obtained within the classical statistical mechanics.

Finally in Sect. XI some concluding remarks are reported, while
the main mathematical properties of the functions
$\exp_{_{\{{\scriptstyle \kappa}\}}}\!(t)$ and
$\ln_{_{\{{\scriptstyle \kappa}\}}}\!(t)$ are collected in
Appendix.

\sect{Deformed sums in special relativity}

In this section we introduce the special relativity in an
alternative way, starting from the well known formulas of
hyperbolic trigonometry (pag. 58 of ref. \cite{Gradshteyn})
\begin{eqnarray}
{\rm arcsinh}\, x = {\rm arccosh}\,\sqrt{1+x^2}={\rm
arctanh}\frac{x}{\sqrt{1+x^2}}  \ \ . \  \label{RII1}
\end{eqnarray}
We pose $x=\kappa q$, being $q$ the dimensionless variable
corresponding to the momentum and $\kappa$ a dimensionless
deformation parameter. After introducing the functions
\begin{eqnarray}
u(q)=\frac{q}{\sqrt{1+\kappa^2q ^2}} \ \ , \label{RII2}
\end{eqnarray}
\begin{eqnarray}
{\cal W}(q)=
\frac{1}{\kappa^2}\sqrt{1+\kappa^2q^2}-\frac{1}{\kappa^2} \ \ ,
\label{RII3}
\end{eqnarray}
corresponding to the dimensionless velocity and kinetic energy
respectively we can  write Eq. (\ref{RII1}) in the form
\begin{eqnarray}
{\rm arcsinh} (\kappa q) = {\rm arccosh}(1+\kappa^2{\cal W})={\rm
arctanh}(\kappa u)  \ \ , \  \label{RII4}
\end{eqnarray}
which in the classical limit $\kappa \rightarrow 0$ reproduces
the dispersion relation $q=\sqrt{2{\cal W}}=u$ of Newton
mechanics. Finally, after introducing the dimensionless total
energy
\begin{eqnarray}
{\cal E}({\cal W})={\cal W}+\frac{1}{\kappa^2} \ \ , \label{RII5}
\end{eqnarray}
one can write Eq. (\ref{RII1}) also in the form
\begin{eqnarray}
{\rm arcsinh} (\kappa q) = {\rm arccosh}(\kappa^2{\cal E})={\rm
arctanh}(\kappa u)  \ \ . \  \label{RII6}
\end{eqnarray}
It is immediate to verify the validity of the following formulas
\begin{eqnarray}
&&q(u)=\frac{u}{\sqrt{1-\kappa^2u^2}} , \label{RII7} \\
&&q({\cal W})=\sqrt{2{\cal W}+\kappa^2{\cal W}^2} \ \ , \label{RII8} \\
&&q({\cal E})=\sqrt{\kappa^2{\cal E}^2-\frac{1}{\kappa^2}} \ \ , \label{RII9} \\
&&u({\cal W})=\frac{\sqrt{2{\cal W}+\kappa^2{\cal
W}^2}}{1+\kappa^2{\cal W}} \ \ , \label{RII10} \\
&&u({\cal E})=\frac{\sqrt{\kappa^2{\cal
E}^2-1/\kappa^2}}{\kappa^2{\cal E}} \ \ , \label{RII11} \\
&&{\cal W}(u)=\frac{1}{\kappa^2\sqrt{1-\kappa^2u^2}}
-\frac{1}{\kappa^2} \ \ , \label{RII12} \\
&&{\cal W}({\cal E})={\cal E}-\frac{1}{\kappa^2} \ \ , \label{RII13} \\
&&{\cal E}(q)=\frac{1}{\kappa^2}\sqrt{1+\kappa^2q^2} \ \ ,
\label{RII14}
\\ &&{\cal E}(u)=\frac{1}{\kappa^2\sqrt{1-\kappa^2u^2}} \ \ . \label{RII15}
\end{eqnarray}

At this point one can define the physical variables velocity $v$,
momentum $p$ and total energy ${E}$ through
\begin{eqnarray}
\frac{v}{u}=\frac{p}{m q}=\sqrt{\frac{E}{m{\cal E}}}=\kappa c=v_*
\ \ , \ \label{RII16}
\end{eqnarray}
and the kinetic energy as $W=E-mc^2$, being $c$ the light speed.
We impose that $\lim_{c\rightarrow \infty,\,\, \kappa\rightarrow
0}\, v_*<\infty$, in order to preserve the validity of the
definitions of the physical variable, also in the classical
limit. After inserting the physical variables in the above
obtained formulas, one recover all the formulas of the one
particle dynamics within the special relativity. For instance,
Eq.s (\ref{RII7}) and (\ref{RII14}) transform
\begin{eqnarray}
&&p=\frac{mv}{\sqrt{1-(v/c)^2}} \ \ , \label{RII17} \\
&&E=\sqrt{m^2c^4+p^2c^2} \ , \label{RII18}
\end{eqnarray}
and so on.

Let us consider in the one-dimension frame $\cal S$ two identical
particles of rest mass $m$. We suppose that the first particle
moves towards right with velocity $u_1$ while the second particle
moves towards left with velocity $u_2$. The momenta of the two
particles are given by $q_1=q\,(u_1)$ and $q_2=q\,(u_2)$
respectively, where $q\,(u)$ is defined through Eq. (\ref{RII7}).
The total energies of the two particles are given by ${\cal
E}_1={\cal E}\,(u_1)$ and ${\cal E}_2={\cal E}\,(u_2)$
respectively, with ${\cal E}\,(u)$ given by Eq. (\ref{RII15}).
Analogously the kinetic energies of the two particles are given
by ${\cal W}_1={\cal W}\,(u_1)$ and ${\cal W}_2={\cal W}\,(u_2)$
respectively, with ${\cal W}\,(u)$ defined by Eq. (\ref{RII12}).

We consider now the same particles in a new frame $\cal S\,'$
which moves at constant speed $u_2$ towards left with respect to
the frame $\cal S$. In this new frame, which is the rest frame of
the second particle, the two particles have velocities given by
$u\,'_1=u_1\stackrel{\kappa}{\oplus}u_2$ and $u\,'_2=0$
respectively, being
\begin{equation}
u_1\stackrel{\kappa}{\oplus}u_2=\frac{u_1+u_2}{1+\kappa^2u_1u_2}
\ \ , \label{RII19}
\end{equation}
the well known relativistic additivity law for the dimensionless
velocities. In the same frame $\cal S\,'$ the particle momenta
are given by $q\,'_1=q\,(u\,'_1)$ and $q\,'_2=0$. Analogously in
$\cal S\,'$, the particle kinetic energies are given by ${\cal
W}\,'_1={\cal W}\,(u\,'_1)$ and ${\cal W}\,'_2=0$, while the
total particle energies results to be ${\cal E}\,'_1={\cal
E}\,(u\,'_1)$ and ${\cal E}\,'_2=1/\kappa^2$.

A very interesting result follows from the relation ${\cal
E}\,'_2=1/\kappa^2$, regarding the physical meaning of the
parameter $\kappa$. It is evident that the quantity $1/\kappa^2$
represents the dimensionless rest energy of the relativistic
particle.

Let us ask now the following questions: is it possible to obtain
the value of the momentum $q\,'_1$ (or of the total energy ${\cal
E}\,'_1$, or of the kinetic energy ${\cal W}\,'_1$) starting
directly from the values of the momenta $q_1$ and $q_2$ (or of the
total energies ${\cal E}\,_1$ and ${\cal E}\,_2$, or of the
kinetic energies ${\cal W}\,_1$ and ${\cal W}\,_2$) in the frame
$\cal S$? The answers to the above questions are affirmative as we
will see in the following.

First we consider the case of the relativistic momentum $q\,'_1$.
Clearly we can write
$q\,'_1=q\,(u\,'_1)=q(u_1\!\!\stackrel{\kappa}{\oplus}\!u_2)$.
Then  following ref. \cite{PRE02} we have
\begin{eqnarray}
q(u_1\stackrel{\kappa}{\oplus}u_2)&=&\frac{u_1\stackrel{\kappa}{\oplus}u_2}
{\sqrt{1-\kappa^2(u_1\stackrel{\kappa}{\oplus}u_2)^2}} \nonumber
\\&=&\frac{u_1\stackrel{\kappa}{\oplus}u_2}
{\sqrt{1-\kappa^2(\frac{u_1+u_2}{1+\kappa^2u_1u_2} )^2}} \nonumber
\\ &=& (u_1\stackrel{\kappa}{\oplus}u_2)\, \frac{1+\kappa^2u_1u_2}
{\sqrt{(1-\kappa^2u_1^2)(1-\kappa^2u_2^2)}} \nonumber \\
&=&\frac{u_1+u_2}
{\sqrt{(1-\kappa^2u_1^2)(1-\kappa^2u_2^2)}}\nonumber
\\ &=& \frac{u_1}
{\sqrt{1-\kappa^2u_1^2}}\sqrt{1+\frac{\kappa^2u_2^2}{1-\kappa^2u_2^2}}
 \nonumber
\\ &+& \frac{u_2}
{\sqrt{1-\kappa^2u_2^2}}\sqrt{1+\frac{\kappa^2u_1^2}{1-\kappa^2u_1^2}}
 \nonumber \\ &=&q_1
\sqrt{1+\kappa^2q_2^2}+q_2 \sqrt{1+\kappa^2q_1^2} \ \ .
\label{RII20}
\end{eqnarray}
Hence we can write
\begin{equation}
{q }\left({ u }_1\stackrel{\kappa}{\oplus}{u }_2\right)={q
}(u_1)\stackrel{\kappa}{\oplus}{q }(u_2) \label{} \ \ ,
\label{RII21}
\end{equation}
being
\begin{eqnarray}
q_1\stackrel{\kappa}{\oplus}q_2 = q_1\sqrt{1\!+\!\kappa^2q_2^2}
+q_2\sqrt{1\!+\!\kappa^2q_1^2} \ \ , \label{RII22}
\end{eqnarray}
the $\kappa$-sum of relativistic momenta introduced in ref.
\cite{PhA01}. In words, the relativistic momentum $q\,'_1$ of the
first particle, in the rest frame of the second particle $\cal
S'$, is the $\kappa$-deformed sum of the momenta $q_1$ and $q_2$
of the two particles in the frame $\cal S$. The $\kappa$-sum of
the relativistic momenta and the relativistic sum of the
velocities are intimately related. They reduce to the standard
sum as the velocity $c$ approaches to infinity (or equivalently
the parameter $\kappa$ approaches to zero). The deformations in
the above sums, in both cases, are relativistic effects and are
originated from the fact that $c$ has a finite value.

We proceed now to calculate the total energy ${\cal E}\,'_1$ of
the first particle in the frame $\cal S\,'$ starting directly from
the values, in the frame $\cal S$, of the total energies ${\cal
E}\,_1$ and ${\cal E}\,_2$. Preliminarily we recall the following
mathematical properties of the inverse hyperbolic trigonometric
functions (pag. 58 of ref. \cite{Gradshteyn})
\begin{eqnarray}
\!\!\!\!\!\!{\rm arcsinh}\, x_1 \!\!\!&\pm&\!\!\!{\rm arcsinh}\,
x_2 \nonumber
\\\!\!\!&=&\!\!\! {\rm arcsinh} \left(\!x_1\sqrt{1\!+\!x_2^2}
\pm x_2\sqrt{1\!+\!x_1^2}\,\right) ,
\label{RII23} \ \ \  \\ \!\!\!\!\!\!{\rm arccosh}\, y_1 \!\!\!&\pm&\!\!\!{\rm arccosh}y_2 \nonumber \\
\!\!\!&=&\!\!\!{\rm arccosh}
\!\left(\!y_1y_2\!\pm\!\sqrt{(y_1^2\!-\!1)(y_2^2\!-
\!1)}\,\right) ,
\label{RII24} \ \ \  \\ \!\!\!\!\!\!{\rm arctanh}\, z_1 \!\!\!&\pm&\!\!\!{\rm arctanh}\, z_2 \nonumber \\
\!\!\!&=&\!\!\! {\rm arctanh}\, \frac{z_1\pm z_2}{1+z_1z_2} \ .
\label{RII25}
\end{eqnarray}
After posing $x=\kappa q$, $y=\kappa^2 {\cal E}$, $z=\kappa u$ and
taking into account Eq. (\ref{RII6}), we are able to define the
$\kappa$-sum of the relativistic total energies as follows
\begin{eqnarray}
{\cal E}_1\stackrel{\kappa}{\oplus}{\cal E}_2 = \kappa^2{\cal
E}_1{\cal E}_2\!+\!\frac{1}{\kappa^2}\sqrt{\big(\kappa^4{\cal
E}_1^2\!-\!1\big)\big(\kappa^4{\cal E}_2^2\!- \!1\big)}\ \ .
\label{RII26}
\end{eqnarray}
After having noticed that
\begin{eqnarray}
{\cal E}(u_1)\stackrel{\kappa}{\oplus}{\cal E }(u_2)&\!\!=\!\!&
\frac{1}{\kappa^2}\frac{1}{\sqrt{1-\kappa^2u^2_1}}\,\stackrel{\kappa}{\oplus}
\frac{1}{\kappa^2}\frac{1}{\sqrt{1-\kappa^2u^2_1}} \nonumber \\
&\!\!=\!\!&
\frac{1}{\kappa^2}\frac{1}{\sqrt{(1-\kappa^2u^2_1)(1-\kappa^2u^2_2)}}
\nonumber \\ &\!\!+\!\!& \frac{1}{\kappa^2}
\sqrt{\left(\frac{1}{1-\kappa^2u^2_1}-1\right)
\left(\frac{1}{1-\kappa^2u^2_2}-1\right)} \nonumber \\
&\!\!=\!\!&
\frac{1}{\kappa^2}\frac{1}{\sqrt{(1-\kappa^2u^2_1)(1-\kappa^2u^2_2)}}
\nonumber \\ &\!\!+\!\!&
\frac{u_1u_2}{\sqrt{(1-\kappa^2u^2_1)(1-\kappa^2u^2_2)}}
\nonumber \\ &\!\!=\!\!&
\frac{1}{\kappa^2}\frac{1+\kappa^2u_1u_2}{\sqrt{(1-\kappa^2u^2_1)(1-\kappa^2u^2_2)}}
\nonumber \\ &\!\!=\!\!& \frac{1}{\kappa^2}
\frac{1+\kappa^2u_1u_2}{\sqrt{(1+\kappa^2u_1u_2)^2-\kappa^2(u_1+u_2)^2}}
\nonumber \\ &\!\!=\!\!& \frac{1}{\kappa^2}
\frac{1}{\sqrt{1-\kappa^2\left(\frac{u_1+u_2}{1+\kappa^2u_1u_2}\right)^2}}
\nonumber \\ &\!\!=\!\!& \frac{1}{\kappa^2}
\frac{1}{\sqrt{1-\kappa^2\left( { u
}_1\stackrel{\kappa}{\oplus}{u }_2 \right)^2}} \nonumber
\\ &\!\!=\!\!& {\cal E}\left({ u }_1\stackrel{\kappa}{\oplus}{u
}_2\right) \ \ , \label{RII27}
\end{eqnarray}
we can conclude that the total energy ${\cal E}\,'_1={\cal
E}\big({ u }_1\stackrel{\kappa}{\oplus}{u }_2\big)$ of the first
particle, in the rest frame of the second particle $\cal S\,'$, is
the $\kappa$-sum, defined through Eq. (\ref{RII26}), of the total
energies ${\cal E}_1$ and ${\cal E}_2$ of the particles in the
frame $\cal S$:
\begin{eqnarray}
{\cal E}\left({ u }_1\stackrel{\kappa}{\oplus}{u }_2\right)={\cal
E}(u_1)\stackrel{\kappa}{\oplus}{\cal E }(u_2) \ \ . \label{RII28}
\end{eqnarray}

The $\kappa$-sum for the relativistic kinetic energies can be
defined starting from
\begin{eqnarray}
{\cal W}_1\stackrel{\kappa}{\oplus}{\cal W}_2 = {\cal
E}_1\stackrel{\kappa}{\oplus}{\cal E}_2 -\frac{1}{\kappa^2} \ \ ,
\label{RII29}
\end{eqnarray}
and assumes the form
\begin{eqnarray}
\!\!\!\!\!\!\!\!{\cal W}_1\stackrel{\kappa}{\oplus}{\cal W}_2
&\!\!=\!\!& {\cal
W}_1\!+\!{\cal W}_2\!+\!\kappa^2{\cal W}_1{\cal W}_2 \nonumber \\
&\!\!+\!\!& \sqrt{{\cal W}_1{\cal W}_2\big(2\!+\!\kappa^2{\cal
W}_1\big)\big(2\!+\!\kappa^2{\cal W}_2\big)}  \ \ . \label{RII30}
\end{eqnarray}
After taking into account Eqs. (\ref{RII12}), (\ref{RII19}) and
(\ref{RII30}) one immediately obtains
\begin{eqnarray}
{\cal W}\left({ u }_1\stackrel{\kappa}{\oplus}{u }_2\right)={\cal
W}(u_1)\stackrel{\kappa}{\oplus}{\cal W }(u_2) \ \ . \label{RII31}
\end{eqnarray}
In words the $\kappa$-sum, defined through Eq. (\ref{RII30}), of
the particle kinetic energies ${\cal W}_1$ and ${\cal W}_2$ in
the frame $\cal S$, gives the kinetic energy of one of the two
particles in the rest frame $\cal S\,'$ of the other particle.

Besides the relationships (\ref{RII21}), (\ref{RII28}) and
(\ref{RII31}), linking the $\kappa$-sums defined through Eqs.
(\ref{RII19}), (\ref{RII22}), (\ref{RII26}) and (\ref{RII30}), one
can easily obtain:

\begin{eqnarray}
&&{q }(u_1)\stackrel{\kappa}{\oplus}{q }(u_2) ={q }\left({ u
}_1\stackrel{\kappa}{\oplus}{u }_2\right) \ \ , \label{RII32} \\
&&{\cal E }(u_1)\stackrel{\kappa}{\oplus}{\cal E}(u_2) ={\cal E
}\left({u}_1\stackrel{\kappa}{\oplus}{u}_2\right) \ \ , \label{RII33} \\
&&{\cal W }(u_1)\stackrel{\kappa}{\oplus}{\cal W}(u_2) ={\cal W
}\left({u}_1\stackrel{\kappa}{\oplus}{u}_2\right) \ \ , \label{RII34} \\
&&{u }(q_1)\stackrel{\kappa}{\oplus}{u}(q_2) ={u
}\left({q}_1\stackrel{\kappa}{\oplus}{q}_2\right) \ \ , \label{RII35} \\
&&{\cal E }(q_1)\stackrel{\kappa}{\oplus}{\cal E}(q_2) ={\cal E
}\left({q}_1\stackrel{\kappa}{\oplus}{q}_2\right) \ \ , \label{RII36} \\
&&{\cal W }(q_1)\stackrel{\kappa}{\oplus}{\cal W}(q_2) ={\cal W
}\left({q}_1\stackrel{\kappa}{\oplus}{q}_2\right) \ \ , \label{RII37} \\
&&q({\cal E}_1)\stackrel{\kappa}{\oplus}q({\cal E}_2) ={q
}\left({\cal E}_1\stackrel{\kappa}{\oplus}{\cal E}_2\right) \ \ , \label{RII38} \\
&&u({\cal E}_1)\stackrel{\kappa}{\oplus}u({\cal E}_2) ={u
}\left({\cal E}_1\stackrel{\kappa}{\oplus}{\cal E}_2\right) \ \ , \label{RII39} \\
&&{\cal W}({\cal E}_1)\stackrel{\kappa}{\oplus}{\cal W}({\cal
E}_2) ={{\cal W}
}\left({\cal E}_1\stackrel{\kappa}{\oplus}{\cal E}_2\right)\ \ , \label{RII40} \\
&&q({\cal W}_1)\stackrel{\kappa}{\oplus}q({\cal W}_2) ={q
}\left({\cal W}_1\stackrel{\kappa}{\oplus}{\cal W}_2\right) \ \ , \label{RII41} \\
&&u({\cal W}_1)\stackrel{\kappa}{\oplus}u({\cal W}_2) ={u
}\left({\cal W}_1\stackrel{\kappa}{\oplus}{\cal W}_2\right) \ \ ,
\label{RII42} \\ &&{\cal E}({\cal
W}_1)\stackrel{\kappa}{\oplus}{\cal E}({\cal W}_2) ={{\cal E}
}\left({\cal W}_1\stackrel{\kappa}{\oplus}{\cal W}_2\right) \ \ .
\label{RII43}
\end{eqnarray}

Let us make some remarks on the meaning of the $\kappa$-sums. We
note that in the classical limit the $\kappa$-sum for the
relativistic kinetic energies given by Eq. (\ref{RII30}) reduces
to the following expression of Galileian relativity
\begin{eqnarray}
{\cal W}_1\stackrel{0}{\oplus}{\cal W}_2 ={\cal W}_1+{\cal
W}_2+2\sqrt{{\cal W}_1{\cal W}_2} \ \ . \label{RII44}
\end{eqnarray}
Then a deformed sum appears also in classical physics when we
change the frame of observation of the particle system. Clearly
this happens only for the kinetic energy. For the classical
momenta and velocities hold the ordinary sum.

We emphasize that the four $\kappa$-sums given by Eqs.
(\ref{RII19}), (\ref{RII22}), (\ref{RII26}) and (\ref{RII30}),
emerge only when we change the frame of observation of the
particles from $\cal S$ to $\cal S'$. More precisely these sums
give the value of $q'_1$ (or $u'_1$, ${\cal E'}_1$, ${\cal
W'}_1$) of the first particle in the rest frame $\cal S'$ of the
second particle, starting directly from the value of $q_1$ (or
$u_1$, ${\cal E}_1$, ${\cal W}_1$) and $q_2$ (or $u_2$, ${\cal
E}_2$, ${\cal W}_2$) of the two particles in the old frame $\cal
S$.

\sect{Minkowski Four-vectors and $\kappa$-exponential}

From the general discussion in the previous section appears clear
that any relativistic formula can be viewed as a one-parameter
deformation of the corresponding classical formula. For instance
the kinetic energy given by the relativistic dispersion relation
\begin{eqnarray}
W=\sqrt{m^2 c^4+p^{2} c^2} -mc^2  \ \ . \label{RIII1}
\end{eqnarray}
can be viewed as the one-parameter relativistic deformation of
the classical parabolic dispersion relation ${ W}_{cl}=p^2/2m$.
In the present section we will show that the relativistic
dispersion relation imposes a one-parameter deformation to the
ordinary exponential function, which presents now as a
mathematical tool of classical physics. The new deformed
exponential, as we will see in the following sections, replaces
the ordinary exponential in the relativistic physics.

We adopt the metric $g^{\mu\nu}={\rm diag}(1,-1,-1,-1)$ in the
Minkowski space and recall that the length of any four-vector is
Lorentz invariant. In particular for the four-momentum
$p^{\mu}=(E/c,\,\bf p \,)$, $p_{\mu}=(E/c,-\bf p \,)$ one obtains
the relativistic dispersion relation
\begin{eqnarray}
 p^{\mu} p_{\mu}=m^2c^2 \ \ , \label{RIII2}
\end{eqnarray}
which in terms of the total energy assumes the form
\begin{eqnarray}
E=\sqrt{m^2 c^4+p^{2} c^2} \ \ . \label{RIII3}
\end{eqnarray}
We can write the latter equation as follows
\begin{eqnarray}
\left(\frac{E}{mc^2}\right)^2 - \left(\frac{p^{}}{m
c^{}}\right)^2=1 \ \ , \label{RIII4}
\end{eqnarray}ù
and then
\begin{eqnarray}
\left(\frac{E}{mc^2}-\frac{p}{m
c}\right)\left(\frac{E}{mc^2}+\frac{p}{m c}\right)=1 \ \ .
\label{RIII5}
\end{eqnarray}
At this point we can eliminate the variable $E$ by employing once
again Eq. (\ref{RIII3}) obtaining
\begin{eqnarray}
\!\!\!\left(\!\sqrt{1\!+\!\left(\frac{p}{m c}\right)^2}-\frac{p}{m
c}\right)\!\left(\!\sqrt{1\!+\!\left(\frac{p}{m
c}\right)^2}+\frac{p}{m c}\right)=1 \ \ . \label{RIII6}
\end{eqnarray}
The latter relationship, after introducing the dimensionless
momentum $q$, defined through Eq.(\ref{RII16}), can be written as
\begin{eqnarray}
\left(\!\sqrt{1\!+\!\kappa^2 q^2}-\kappa
q\right)\!\left(\!\sqrt{1\!+\!\kappa^2q^2}+\kappa q\right)=1 \ \
, \label{RIII7}
\end{eqnarray}
or equivalently
\begin{eqnarray}
\left(\!\sqrt{1\!+\!\kappa^2 q^2}-\kappa
q\right)^{1/\kappa}\!\left(\!\sqrt{1\!+\!\kappa^2 q^2}+\kappa
q\right)^{1/\kappa}\!\!=1  \ \ . \label{RIII8}
\end{eqnarray}

We remark that Eq. (\ref{RIII8}) follows directly from the
dispersion relation. Interestingly in the classical limit
${\kappa \rightarrow 0}$ Eq. (\ref{RIII8}) reduces to
$\exp(-q)\,\exp(q)=1$ while the dispersion relation becomes the
classical one $W=p^2/2m$. In this way we obtain a direct link
between the dispersion relation of free classical particles and
the ordinary exponential function.

In the light of the above result, we reconsider Eq. (\ref{RIII8})
which, after introducing the function
\begin{eqnarray}
\exp_{_{\{{\scriptstyle \kappa}\}}}\!\left(q
\right)=\left(\!\sqrt{1\!+\!\kappa^2 q^2} +\kappa
q\right)^{1/\kappa}  \ \ , \label{RIII9}
\end{eqnarray}
writes in the form
\begin{eqnarray}
\exp_{_{\{{\scriptstyle \kappa}\}}}\!\left( -\,q \right)\,\,
\exp_{_{\{{\scriptstyle \kappa}\}}}\!\left( q \right) =1  \ \ .
\label{RIII10}
\end{eqnarray}
The function $\exp_{_{\{{\scriptstyle \kappa}\}}}\!\left(q
\right)$, reproducing the exponential function,
$\exp(q)=\exp_{_{\{{\scriptstyle 0}\}}}\!\left(q \right)$ in the
classical limit $\kappa \rightarrow 0$, represents a
one-parameter relativistic generalization of the ordinary
exponential.

\sect{Lorentz invariant integration and $\kappa$-exponential}

In this section we present a second path which unambiguously
conducts to the $\exp_{_{\{{\scriptstyle \kappa}\}}}\!\left(q
\right)$ function.  Let us consider the following integral in the
three dimension momentum space, within the classical physics
framework
\begin{eqnarray}
I_{cl}=\int d^3q \,F \ \ , \label{RIV1}
\end{eqnarray}
being $F$ an arbitrary quantity. Whenever $F$ depends only on
$q=|{\bf q}|$, the above integral can be reduced to the following
one dimension integral
\begin{eqnarray}
I_{cl}=\int_0^{\infty} dq\,\,4\pi\, q^2 \,F \ \ . \label{RIV2}
\end{eqnarray}
After introducing the physical variable ${\bf
p}=mv_*^{\infty}\,{\bf q}$ the integral (\ref{RIV1}) transforms
\begin{eqnarray}
I_{cl}=\int \frac{d^3p}{(mv_{*}^\infty)^{\,3}} \,F  \ \ .
\label{RIV3}
\end{eqnarray}

In the framework of a relativistic theory it is well known that
the integral (\ref{RIV3}) must be replaced by a Lorentz invariant
integral $I_{cl} \rightarrow I_{rel}$, being
\begin{eqnarray}
I_{rel}=A \int d^4p
\,\,\,\theta(p_0)\,\delta(p^{\mu}p_{\mu}-m^2c^2) \, \,F \ \ ,
\label{RIV4}
\end{eqnarray}
and $A$ a constant. Now we can introduce the variable ${\bf
q}={\bf p}\,/mv_{\,*}$ and take into account that
$p^{\mu}=(E/c,\, \bf p \,)$ and $E=\sqrt{m^2 c^4+p^{2} c^2}$.
After properly choosing the value of $A$,  the four-dimension
integral (\ref{RIV4}) can be easily reduced to the following
three dimension integral
\begin{eqnarray}
I_{rel}=\int \frac{d^3 q}{\sqrt{1+\kappa^2 q^2 }}\,F \ \ .
\label{RIV5}
\end{eqnarray}
We remark that in (\ref{RIV4}) the integration element $d^4p$ is a
scalar because the Jacobian of the Lorentz transformation is
equal to unity. Then $I_{rel}$ transforms as $F$. For this reason
in (\ref{RIV5}) the integration element $d^3 q/\sqrt{1+\kappa^2
q^2 }$ is a scalar. Finally we can reduce (\ref{RIV5}) in the
following one dimension integral
\begin{eqnarray}
I_{rel}=\int_0^{\infty} \frac{d q}{\sqrt{1+\kappa^2 q^2 }}\,4\pi\,
q^2 \,F \ \ . \label{RIV6}
\end{eqnarray}
It is important to note that in classical limit $\kappa
\rightarrow 0$, Eqs. (\ref{RIV5}) and (\ref{RIV6}) reproduce the
corresponding classical ones given by (\ref{RIV1}) and
(\ref{RIV2}) respectively.

We focus now our attention to the classical and relativistic
expression of the one dimension integrals given by (\ref{RIV2})
and (\ref{RIV6}) respectively. One immediately observes that the
relativistic integral is obtained directly from the classical
one, by making the substitution $d q \rightarrow
d_{{\scriptscriptstyle\{}\kappa{\scriptscriptstyle\}}}q$, being
\begin{eqnarray}
d_{{\scriptscriptstyle\{}\kappa{\scriptscriptstyle\}}}q=\frac{d
q}{\sqrt{1+\kappa^2 q^2 }} \ \ , \label{RIV7}
\end{eqnarray}
the $\kappa$-differential. The relativistic one dimension integral
can be written in the form
\begin{eqnarray}
I_{rel}=\int_0^{\infty}d_{{\scriptscriptstyle\{}\kappa{\scriptscriptstyle\}}}q
\,\,4\pi\, q^2 \,F \ \ . \label{RIV8}
\end{eqnarray}
Clearly the $\kappa$-integral $\int
dq_{{\scriptscriptstyle\{}\kappa{\scriptscriptstyle\}}} $ is
originated from the fact that the light speed has a finite value.

As a working example, involving explicitly the
$\kappa$-integration, we consider the definition of the kinetic
energy. In classical mechanics the kinetic energy, using
dimensionless variables, is defined as
\begin{eqnarray}
{\cal W}_{cl}(q)=\int_0^q  dq \,\, q \ \ . \label{RIV9}
\end{eqnarray}
It is straightforward to verify that the extension of the above
definition in special relativity is given by
\begin{eqnarray}
{\cal W}(q)=\int_0^q
d_{{\scriptscriptstyle\{}\kappa{\scriptscriptstyle\}}}q \,\, q \
\ . \label{RIV10}
\end{eqnarray}
Then, the replacement of the ordinary integration with by the
$\kappa$-integration, in the classical definition (\ref{RIV9}),
permits us to recover the relativistic expression (\ref{RII3}) of
the kinetic energy.

Let us consider now the $\kappa$-derivative related to the above
defined $\kappa$-integration
\begin{eqnarray}
\frac{d}{d_{{\scriptscriptstyle\{}\kappa{\scriptscriptstyle\}}}q}=
\sqrt{1+\kappa^2 q^2 }\,\,\frac{d}{d q} \ \ . \label{RIV11}
\end{eqnarray}
Eq. (\ref{RIV10}) linking ${\cal W}$ and $q$ can be written now
also in the following differential form
\begin{eqnarray}
\frac{d}{d_{{\scriptscriptstyle\{}\kappa{\scriptscriptstyle\}}}q}
\,\,{\cal W}(q) = q \ \ , \label{RIV12}
\end{eqnarray}
which, after integration with the condition ${\cal W}(0)=0$,
yields the relativistic expression of the kinetic energy
(\ref{RII3}). In the  $\kappa \rightarrow 0$ limit the above
differential equation reduces to the classical one $(d/dq){\cal
W}(q) = q$.

Taking into account that the ordinary exponential results to be
eigenfunction of the ordinary derivative, namely
$(d/dq)\exp(q)=\exp(q)$, the question to determine the
eigenfunction of the $\kappa$-derivative,  naturally arises.
After some simple calculations one obtains
\begin{eqnarray}
\frac{d}{d_{{\scriptscriptstyle\{}\kappa{\scriptscriptstyle\}}}q}
\exp_{_{\{{\scriptstyle \kappa}\}}}\!\left(q
\right)=\exp_{_{\{{\scriptstyle \kappa}\}}}\!\left(q \right) \ \
, \label{RIV13}
\end{eqnarray}
so that the $\kappa$-exponential arises as eigenfunction of the
$\kappa$-derivative.

\sect{Relativistic sums and $\kappa$-diferential}

In this section we explain the physical origin of the mechanism
enforcing the deformation in the $\kappa$-differential. We
consider two identical relativistic particles with rest mass $m$
and velocities ${\bf v}_1$ and ${\bf v}_2$ in the three-dimension
frame ${\cal S}$. The modulus of the relative velocity $V$ of the
particles depends on ${\bf v}_1$ and ${\bf v}_2$ and is given by
(page 20 of ref. \cite{Cercignani})
\begin{eqnarray}
V({\bf v}_1,{\bf v}_2)= \sqrt{\left({\bf v}_1\ominus {\bf
v}_2\right)^2-\frac{1}{c^2}\left(\frac{{\bf v}_1 \times{\bf
v}_2}{1-{\bf v}_1{\bf v}_2/c^2}\right)^2} , \label{RV1}
\end{eqnarray}
being
\begin{equation}
{\bf v}_1\ominus {\bf v}_2=\frac{{\bf v}_1-{\bf v}_2}{1-{\bf
v}_1{\bf v}_2/c^2} \ \ . \label{RV2}
\end{equation}

We perform now the calculation of the modulus of relative
momentum related to the modulus of the relative velocity of the
two particles according to the relativistic formula
\begin{equation}
P(V)=\frac{m V}{\sqrt{1-V^2/c^2}{}} \ \ . \label{RV3}
\end{equation}
Clearly results that $P(V)=P({\bf v}_1,{\bf v}_2)$ and after
taking into account that
\begin{equation}
{\bf v}=\frac{{\bf p}/m}{\sqrt{1+{\bf p}^2/m^2c^2}{}} \ \ ,
\label{RV4}
\end{equation}
we can conclude that $P(V)=P({\bf p}_1,{\bf p}_1)$. At this point
we introduce the dimensionless variables ${\bf q}={\bf p}/mv_{*}$
and $Q=P/mv_{*}$. After tedious but straightforward calculation,
one arrives to the following expression for the modulus of the
dimensionless relative momentum
\begin{eqnarray}
{\rm Q}({\bf q}_1,{\bf q}_2)= \sqrt{\frac{\left({\bf q}_1
\stackrel{\kappa}{\ominus}{\bf q}_2\right)^2-\kappa^2\left({\bf
q}_1 \times{\bf q}_2\right)^2}{1-\kappa^4\left({\bf q}_1
\times{\bf q}_2\right)^2} } \ \ , \label{RV5}
\end{eqnarray}
being
\begin{eqnarray}
{\bf q}_1 \stackrel{\kappa}{\ominus}{\bf q}_2={\bf
q}_1\sqrt{1+\kappa^2{\bf q}_2^2}-{\bf q}_2\sqrt{1+\kappa^2{\bf
q}_1^2} \ \ , \label{RV6}
\end{eqnarray}
the $\kappa$-difference introduced in ref. \cite{PhA01}. Clearly
${\rm Q}({\bf q}_1,{\bf q}_2)$ represents the modulus of momentum
of the first (second) particle in the rest frame of second (first)
particle.

We suppose now that the two particles in the frame $S$ have
momenta ${\bf q}_1={\bf q}+d{\bf q}$ and ${\bf q}_2={\bf q}$
respectively and calculate the value of {\rm Q}({\bf q}+d{\bf
q},\,{\bf q}). One easily obtains
\begin{eqnarray}
{\rm Q}({\bf q}+d{\bf q},{\bf
q})=d_{{\scriptscriptstyle\{}\kappa{\scriptscriptstyle\}}}q \ \ ,
\label{RV7}
\end{eqnarray}
being $q=|{\bf q}|$ and $
d_{{\scriptscriptstyle\{}\kappa{\scriptscriptstyle\}}}q$ given by
Eq. (\ref{RIV7}).

The physical meaning of $\kappa$-differential
$d_{{\scriptscriptstyle\{}\kappa{\scriptscriptstyle\}}}q$
immediately follows.  The modulus of the infinitesimal difference
 of the two particle momenta $dq$, in the frame $S$, becomes
$d_{{\scriptscriptstyle\{}\kappa{\scriptscriptstyle\}}}q$, if this
difference is observed in the rest frame of one of the two
particles. The meaning of the $\kappa$-derivative follows also
readily. If $d/dq$ is the derivative in the frame $S$, the
deformed derivative
$d/d_{{\scriptscriptstyle\{}\kappa{\scriptscriptstyle\}}}q$
represents an ordinary derivative in the rest frame of one of the
two particles.

\sect{Relativistic sums and $\kappa$-deformed logarithm and
exponential}

The ordinary logarithm $h(x)=\ln(x)$ is the only existing function
(unless a multiplicative constant) which results to be solution of
the function equation $h(x_1x_2)=h(x_1)+h(x_2)$. Let us consider
now the generalization of this equation within the special
relativity obtained by substituting the ordinary sum by the
$\kappa$-sum of the dimensionless relativistic momenta
\begin{eqnarray}
h(x_1x_2)= h(x_1)\stackrel{\kappa}{\oplus} h(x_2) \ \ .
\label{RVI1}
\end{eqnarray}
We proceed by solving this equation, which assumes the explicit
form
\begin{eqnarray}
h(x_1x_2)&=& h(x_1)\,\sqrt{1+\kappa^2\,h(x_2)\,^2} \nonumber \\
&+& h(x_2)\,\sqrt{1+\kappa^2\,h(x_1)\,^2} \ \ \ . \label{RVI2}
\end{eqnarray}
After performing the substitution $h(x)=\kappa^{-1} \sinh \kappa
g(x)$ we obtain that the auxiliary function $g(x)$ obeys the
equation $g(x_1x_2)=g(x_1)+g(x_2)$, and then is given by
$g(x)=A\ln x$. The unknown function $h(x)$ becomes
\begin{eqnarray}
h(x)=\frac{\sinh ( \kappa \ln x)}{\kappa} \ \ , \label{RVI3}
\end{eqnarray}
where we have set $A=1$ in order to recover, in the limit
$\kappa\rightarrow 0$, the classical solution $h(x)=\ln(x)$. The
function given by Eq. (\ref{RVI3}) in the following is denoted by
$\ln_{_{\{{\scriptstyle \kappa}\}}}\! (x)$ and can be written
also in the form
\begin{eqnarray} \ln_{_{\{{\scriptstyle \kappa}\}}}\!
(x)=\frac{x^\kappa-x^{-\kappa}}{2\kappa} \ \ . \label{RVI4}
\end{eqnarray}
Holding $\exp_{_{\{{\scriptstyle
\kappa}\}}}\!\!\big(\ln_{_{\{{\scriptstyle \kappa}\}}}
\!(x)\big)=\ln_{_{\{{\scriptstyle
\kappa}\}}}\!\!\big(\exp_{_{\{{\scriptstyle \kappa}\}}}\!
(x)\big)=x$ it results that $\ln_{_{\{{\scriptstyle \kappa}\}}}\!
(x)$ is the inverse function of $\exp_{_{\{{\scriptstyle
\kappa}\}}} \!(x)$ and thus can be viewed as a generalization of
the ordinary logarithm, $\ln(x)=\ln_{_{\{{\scriptstyle 0}\}}}\!
(x)$, in the framework of special relativity.

It is straightforward to obtain the $\exp_{_{\{{\scriptstyle
\kappa}\}}} \!(x)$, starting directly from $\kappa$-sum of the
dimensionless relativistic momenta, as solution of the functional
equation
\begin{eqnarray}
\exp_{_{\{{\scriptstyle \kappa}\}}}
\!(x_1)\exp_{_{\{{\scriptstyle \kappa}\}}} \!(x_2)=
\exp_{_{\{{\scriptstyle \kappa}\}}}
\!(x_1\stackrel{\kappa}{\oplus}x_2) \ \ , \label{RVI5}
\end{eqnarray}
which in the classical limit reduces to the functional equation
$\exp(x_1)\exp(x_2)=\exp(x_1+x_2)$ defining the ordinary
exponential.

In the Appendix, the main mathematical properties of the
functions $\kappa$-exponential and $\kappa$-logarithm are
reported.

\sect{Maximum entropy principle}

We start by recalling some elements of classical statistical
mechanics. We consider the Boltzmann-Gibbs-Shannon entropy
\begin{eqnarray}
S=-\int d^3p \,\,\,f\,(p)\, \ln(f\,(p)) \ \ , \label{RVII1}
\end{eqnarray}
and the constraints fuctional
\begin{equation}
C={\rm a_{_0}}\!\!\left[\,\int\!\! d^3p \,\, f(p)\!-\!1
\right]\!+ {\bf a}\!\left[{\bf M}  \!-\!\!\int\!\! d^3p \, \,
\,{\bf g}(p)\,f(p)\right], \label{RVII2}
\end{equation}
where the constants ${\rm a_{_0}}$  and ${\bf a}=\{{\rm
a_{_1},\,a_{_2},...,\,a_l}\}$ are the $l+1$ Lagrange multipliers,
while the $l$ moments ${\bf M}=\{{\rm
M_{_1},\,M_{_2},...,\,M_l}\}$ are the mean values of the $l$
functions ${\bf g}(p)=\{{\rm g_{_1}}(p),\,{\rm
g_{_2}}(p),...,\,{\rm g_l}(p)\}$. The variational equation
\begin{eqnarray}
\frac{\delta}{\delta f}\, (S+C)=0 \ \ , \label{RVII3}
\end{eqnarray}
means that the entropy $S(p)$ is maximized under the constraints
imposing the conservation of the norm of $f(p)$ and the a priori
knowledge of the values  of the $l$ moments $M_j$ of $f(p)$,
namely
\begin{eqnarray}
\int d^3p \, f(p)=1 \ \ , \ \ M_j=\int d^3p \, \,\,{\rm
g}_j(p)\,f(p) \ \ . \label{RVII4}
\end{eqnarray}
The maximum entropy principle  expressed by the variational
equation (\ref{RVII3}) yields the following statistical
distribution
\begin{eqnarray}
f(p)=\frac{1}{{\rm e}} \exp\big( -{\bf a \cdot g }(p)+{\rm
a}_{_0}\big) \ \ . \label{RVII5}
\end{eqnarray}
In a classical many body particle system the collisional
invariants are essentially the particle number and the kinetic
energy $W(p)=p^2/2m$. Then we need to involve  two Lagrange
multipliers ${\rm a_{_0}}=\mu/k_B T$, ${\rm a_{_1}}=1/k_B T$ and
one moment ${\rm M_1}=\int d^3 p \,\, {\rm g_1}(p)\,f(p)=\int d^3
p \,\,W(p)\,f(p)=<\!\!W\!\!>$. In this way the distribution
function given by Eq. (\ref{RVII5}) reduces to the
Maxwell-Boltzmann one of classical statistical mechanics
\begin{eqnarray}
f(p)=\frac{1}{{\rm e}} \exp\left(- \frac{W(p)-\mu}{k_B T}\right)
\ \ . \label{RVII6}
\end{eqnarray}

Let us consider now the following generalized trace form entropy
\begin{eqnarray}
S=-\int d^3p \,\,\, f\,(p)\, \Lambda(f\,(p)) \ \ , \label{RVII7}
\end{eqnarray}
being $\Lambda(f)$ a function generalizing the ordinary logarithm.
The variational equation (\ref{RVII3}) with $S$ and $C$ given by
(\ref{RVII7}) and (\ref{RVII2}) respectively, produces the
distribution function $f=f(p)$ defined through
\begin{eqnarray}
\frac{\partial}{\partial f}\,f\,\Lambda(f)=-{\bf a \cdot g
}(p)+{\rm a}_{_0} \ \ . \label{RVII8}
\end{eqnarray}

The maximum entropy principle, in the case of the ordinary
statistical mechanics,  relates the Boltzmann-Gibbs-Shanon entropy
and the Maxwell-Boltzmann distribution by means of a twofold link.
Firstly, the distribution is obtained by maximizing the entropy
under proper constraits. Secondly, both the entropy and the
distribution are expressed in terms of the same function in
direct and inverse form. Indeed, the Maxwell-Boltzmann
distribution is given in terms of the ordinary exponential while
the Boltzmann-Gibbs-Shanon entropy is defined starting from the
ordinary logarithm. Clearly, this is a very strong interpretation
of the maximum entropy principle.  In generalized statistical
theories, customary, is used a more weak interpretation of the
maximum entropy principle by relaxing the second link.

Our next task is to use the strong interpretation of the maximum
entropy principle holding for the ordinary statistical mechanics
in order to generalize minimally it. The requirement of the
twofold link between entropy and distribution function also in
generalized statistical theories imposes that the function
$\Lambda(f)$ is a solution of the following
functional-differential equation
\begin{eqnarray}
\frac{\partial }{\partial f} \,f \,\Lambda(f) = \lambda\,\,
\Lambda (f/\alpha)+ \eta \ \ , \label{RVII9}
\end{eqnarray}
being $\{ \alpha, \lambda, \eta \}$ three real constants. Indeed,
the distribution given by (\ref{RVII8}) assumes now the form
\begin{eqnarray}
f(p)=\alpha \,\Lambda^{-1}\!\left(-\,\frac{{\bf a \cdot g
}(p)-{\rm a}_{_0}+\eta}{\lambda}\right) \ \ , \label{RVII10}
\end{eqnarray}
being $\Lambda^{-1}(x)$ the inverse function of $\Lambda(x)$.
Note that the constants $\lambda$ and $\eta$ introduce a scaling
in the Langrange multipliers. Clearly in the classical limit it
must result: $\Lambda(x)\rightarrow \ln(x)$,
$\Lambda^{-1}(x)\rightarrow \exp(x)$ and $\{ \alpha, \lambda,
\eta \}\rightarrow\{1/{\rm e}, 1, 0 \}$.

The problem related to the determination of the generalized
logarithm $\Lambda(f)$ satisfying Eq. (\ref{RVII9}) and the
conditions $\Lambda(1)=0$, $\Lambda'(1)=1$ and $0\Lambda(0)=0$,
can be solved easily. Beside the ordinary logarithm, Eq.
(\ref{RVII9}) admits other new solutions. The general solution of
this equation defines  a three-parameter family of generalized
logarithms indicated by $\Lambda(f)=\ln_{_{\{{\scriptstyle
\kappa,\,r,\, \varsigma}\}}}\!(f)$ being
\begin{eqnarray}
\ln_{_{\{{\scriptstyle \kappa,\,r,\, \varsigma}\}}}\!(f)=
\frac{\varsigma^{\kappa}\, f^{r+\kappa} -\varsigma^{-\kappa}\,
f^{r-\kappa} -\varsigma^{\kappa} +\varsigma^{-\kappa}
}{(\kappa+r)\varsigma^{\kappa}+ (\kappa-r)\varsigma^{-\kappa}} \
\ , \ \ \ \ \ \label{RVII11}
\end{eqnarray}
while $\{ \kappa, r, \varsigma \}$ are three free parameters. The
constants $\alpha$, $\lambda$ and $\eta$ are given by
\begin{eqnarray}
&&\alpha=\left(\frac{1+r-\kappa}{1+r+\kappa}\right)^{\frac{1}{2\,\kappa}}
\ \ , \label{RVII12} \\ &&
\lambda=\frac{\big(1+r-\kappa\big)^{\frac{r+\kappa}{2\,\kappa}}}
{\big(1+r+\kappa\big)^{\frac{r-\kappa}{2\,\kappa}}}
\ \ , \label{RVII13} \\
&&\eta=(\lambda-1)\,\frac{\varsigma^{\kappa}-\varsigma^{-\kappa}}
{(\kappa+r)\varsigma^{\kappa}+ (\kappa-r)\varsigma^{-\kappa}} \ \
. \label{RVII14}
\end{eqnarray}

Some specific choice for the parameters $\{ \kappa, r, \varsigma
\}$, generating particular expressions for the function
$\Lambda(f)$ and for its inverse function
$\Lambda^{-1}(x)=\exp_{_{\{{\scriptstyle \kappa,\,r,\,
\varsigma}\}}}\!(x)$, have been already considered in mathematical
statistics, information theory and statistical mechanics
\cite{PRE05LKS}. As stressed previously, in ref.s
\cite{PhA01,PRE02,PRE05LKS,Abe,Naudts,Chavanis}, it is shown that
it is possible to develop selfconsistent statistical theories
starting from any generalized entropy satisfying some standard
requirements. Fortunately, all these requirements are satisfied by
the entropy defined through (\ref{RVII7}) with
$\Lambda(f)=\ln_{_{\{{\scriptstyle \kappa,\,r,\,
\varsigma}\}}}\!(f)$.

Let us focus now our attention to the  solution (\ref{RVII11})
corresponding to the particular choice $\{ \kappa, r=0,
\varsigma=1 \}$ \cite{PhA01}. It is easy to verify that results
$\ln_{_{\{{\scriptstyle \kappa,\,0,\, 1}\}}}\!(x)
=\ln_{_{\{{\scriptstyle \kappa}\}}}(x)$ and
$\exp_{_{\{{\scriptstyle \kappa,\,0,\,
1}\}}}\!(x)=\exp_{_{\{{\scriptstyle \kappa}\}}}(x)$ with
\begin{eqnarray}
&&\ln_{_{\{{\scriptstyle \kappa}\}}}(x)=
\frac{x^{\kappa}-x^{-\kappa}}{2 \kappa} \ \ , \label{RVII15} \\
&&\exp_{_{\{{\scriptstyle \kappa}\}}}(x)=\left(\sqrt{1+\kappa^2
x^2}+\kappa x \right)^{1/\kappa} \ \ . \label{RVII16}
\end{eqnarray}
Thus the $\kappa$-exponential and $\kappa$-logarithm, emerging in
the special relativity in the place of the ordinary exponential
and logarithm, reappear also in the statistical theory through the
maximum entropy principle. It is immediate to verify that the
reality and positivity of the parameters $\alpha$ and $\lambda$
impose that $-1<\kappa<1$.

Furthermore, in the case $\{ \kappa, 0, \varsigma \}$, Eq.
(\ref{RVII11}) produces a two parameter generalized logarithm and
exponential which results to be the scaled $\kappa$-logarithm and
the scaled $\kappa$-exponential respectively, defined as

\begin{eqnarray}
&&\!\!\!\!\!\!\!\!\!\!\!\!\!\!\!\!\!\!\ln_{_{\{{\scriptstyle
\kappa,\, \varsigma}\}}} (x)= \frac{\ln_{_{\{{\scriptstyle
\kappa}\}}}\! (\varsigma x)-\ln_{_{\{{\scriptstyle \kappa}\}}}
\!\varsigma}{\sqrt{1+\kappa^2\, \ln^{\,2}_{_{\{{\scriptstyle
\kappa}\}}} \!\varsigma }} \ , \label{RVII17}
\\
&&\!\!\!\!\!\!\!\!\!\!\!\!\!\!\!\!\!\!\exp_{_{\{{\scriptstyle
\kappa, \, \varsigma}\}}}
\!(x)\!=\!\frac{1}{\varsigma}\exp_{_{\{{\scriptstyle \kappa}\}}}
\!\! \left(\! \,x\,\sqrt{1\!+\!\kappa^2
\ln^{\,2}_{_{\{{\scriptstyle \kappa}\}}} \!\varsigma } \!+\!
\ln_{_{\{{\scriptstyle \kappa}\}}} \!\varsigma \!\right). \ \
\label{RVII18}
\end{eqnarray}

We remark that the functions $\exp_{_{\{{\scriptstyle
\kappa}\}}}\!(x)$ and $\ln_{_{\{{\scriptstyle \kappa}\}}}\!(x)$
generalizing the ordinary exponential and logarithm respectively,
emerge both in special relativity and independently in many body
physics, through the maximum entropy principle, the cornerstone of
statistical mechanics. This interesting fact suggests us to
consider with a particular attention the statistical theories,
enforced by the entropies related to the generalized logarithms
defined through (\ref{RVII15}) and (\ref{RVII17}).

In the next section we consider the statistical theory involving
the generalized logarithm (\ref{RVII15}).

\sect{Generalized Distribution function}

Let us consider the entropy $S=-<\!\ln_{_{\{{\scriptstyle
\kappa}\}}}\!f\!>$ related to the generalized logarithm defined
through (\ref{RVII15}). Here the distribution function $f=f(x,p)$
depends on the four-vectors position $x$ and momentum $p$. The
explicit form of this entropy is given by
\begin{eqnarray}
S=-\int d^3p \,\,\, f\, \ln_{_{\{{\scriptstyle \kappa}\}}}(f) \ \
, \label{RVIII1}
\end{eqnarray}
and after maximized under the constrains (\ref{RVII2}) by solving
the variational equation
\begin{eqnarray}
\frac{\delta }{\delta f} (S+C)=0 \ \ , \label{RVIII2}
\end{eqnarray}
yields the distribution function
\begin{eqnarray}
f=\alpha \exp_{_{\{{\scriptstyle \kappa}\}}}\!\!\left(-\, \frac{
{\bf a \cdot g }-{\rm a}_{_0}}{\lambda}\right) \ \ ,
\label{RVIII3}
\end{eqnarray}
with ${\bf g}={\bf g}(x, p)$ and
\begin{eqnarray}
&&\alpha=\left(\frac{1-\kappa}{1+\kappa}\right)^{1/2\kappa} \ \ , \label{RVIII4} \\
&&\lambda=\sqrt{1-\kappa^2} \ \ , \label{RVIII5}
\end{eqnarray}
being $|\kappa|<1$. Before to proceed further we explain briefly
the meaning of the constants $\alpha$ and $\lambda$, related by
$1/\lambda=\ln_{_{\{{\scriptstyle \kappa}\}}}\!(1/\alpha)$. We
write the entropy in the form $S=\int_{\cal R} d^3p \,\,\,
\sigma\,(f)$ where $\sigma\,(f)=- f\, \ln_{_{\{{\scriptstyle
\kappa}\}}}(f)$ is the entropy density.  It is easy to verify that
for $f=\alpha$ the entropy density assumes its maximum value,
which is given by $\sigma_{max}=\alpha / \lambda$.

We focus now our attention on the form of ${\bf g}={\bf g}(x,
p)$.  In ref. \cite{Degroot} it is shown that for a relativistic
many body system, in presence of external electromagnetic fields,
the more general microscopic invariant has the form
$\left(p^{\nu}+q A^{\nu}\!/c \right)\,U_{\nu}+$ constant, being
$U_{\nu}$  the hydrodynamic four-vector velocity with
$U^{\nu}U_{\nu}=c^2$. Then we can pose $g_1=\left(p^{\nu}+q
A^{\nu}\!/c \right)\,U_{\nu}-mc^2$,  ${\rm a}_1=1/k_BT$ and ${\rm
a}_0=\mu/k_BT$. The distribution (\ref{RVIII3}), in the case of
relativistic statistical systems, assumes the form
\begin{equation}
f= \alpha\exp_{_{\{{\scriptstyle
\kappa}\}}}\!\!\bigg(-\frac{\left(p^{\nu}+e A^{\nu}\!/c
\right)\,U_{\nu}-mc^2-\mu }{\lambda\,k_{_{B}}T}\bigg) \ \ ,
\label{RVIII6}
\end{equation}
and results to be quite different with respect the Juttner
distribution of the ordinary relativistic statistical mechanics,
where in place of $\kappa$-exponential appears the ordinary
exponential \cite{Degroot}. The distribution (\ref{RVIII6}), in
the global rest frame where $U^{\nu}=(c,0,0,0)$ and in absence of
external forces ($A^{\nu}=0$), simplifies as
\begin{equation}
f=\alpha \exp_{_{\{{\scriptstyle \kappa}\}}}\!\!\left(-\,
\frac{W-\mu}{\lambda\,k_{_{B}}T}\right) \ \ , \label{RVIII7}
\end{equation}
being $W$ the relativistic kinetic energy
\begin{equation}
W=\sqrt{m^2c^4+{\mathbf p}^2c^2}-mc^2 \ \ , \label{RVIII8}
\end{equation}
and in the $\kappa \rightarrow 0$ limit reduces to the classical
distribution (\ref{RVII6}). It is remarkable that the above
distribution (\ref{RVIII7}) presents a power law asymptotic
behavior, namely
\begin{equation}
f \propto W^{-1/\kappa}  \ \ , \ \ (W\rightarrow \infty) \ \ ,
\label{RVIII9}
\end{equation}
in contrast with the ordinary relativistic distribution
(Maxwell-Boltzmann distribution where the energy is given by its
relativistic expression) which decays exponentially.

\sect{Generalized Statistics and kinetics}

In this section we report the main features of statistical theory,
based on the entropy (\ref{RVIII1}), some of which are already
known in the literature.

{\it Equiprobability:} The non negative entropy (\ref{RVIII1})
achieves its maximum value at equiprobability, $f(p)=1/\Omega$ for
$\forall p$, and this value is $S=\ln_{_{\{{\scriptstyle
\kappa}\}}}\!\!\Omega$.

{\it Thermodynamic stability:} The entropy (\ref{RVIII1}) is
concave
\begin{equation}
S[tf_1+(1-t)f_2]\geq tS[f_1]+(1-t)S[f_2] \ \ , \label{RIX1}
\end{equation}
so that the system, in thermodynamic equilibrium, is stable
\cite{PRE02}.

{\it Lesche stability:} The Lesche stability condition is
satisfied by any physically meaningful quantity. In particular it
is a necessary condition to experimentally detect a physical
observable. The entropy given by (\ref{RVIII1}), being a physical
quantity depending on a probability distribution, should exhibit
a small relative error
\begin{equation}
R=\left|\frac{S[f]-S[f']}{{\rm sup}(S[f])}\right|\ \ ,
\label{RIX2}
\end{equation}
with respect to small changes of the probability distribution
\begin{equation}
D=||f-f'|| \ \ . \label{RIX3}
\end{equation}
Mathematically this implies that: For any $\varepsilon>0$ there
exists $\delta>0$ such that $R\leq \varepsilon$ holds for all
distribution functions satisfying $D\leq \delta$. The Lesche
stability condition holds for the Boltzmann-Gibbs-Shannon entropy
and is proved also for the entropy (\ref{RVIII1}) in ref.
\cite{PhA04}. In the thermodynamic limit it is obtained
$\delta=\varepsilon^{1/(1-|\kappa|)}$.

{\it Relativistic kinetic equation:} The statistical distribution
(\ref{RVIII7}) can be viewed as stationary case of a distribution
function $f=f(x,p)$, describing a relativistic many body system,
governed by the generalized Boltzmann evolution equation
\cite{PRE02}
\begin{eqnarray}
\!\!\!\!\!\!p^{\,\nu}\partial_{\nu}f-m F^{\nu}\frac{\partial
f}{\partial p^{\,\nu}}= \int
\frac{d^3p'}{{p'}^{0}}\frac{d^3p_1}{p_1^{\,0}}\frac{d^3p'_1}{{p'}_{\!\!1}^{0}}
\,\,G \,\,&& \nonumber \\
\times
\left[\,a\,(f'\otimes\mbox{\raisebox{-2mm}{\hspace{-3.3mm}$\scriptstyle
\kappa$}} \hspace{1mm}f'_1)-
a\,(f\otimes\mbox{\raisebox{-2mm}{\hspace{-3.3mm}$\scriptstyle
\kappa$}} \hspace{1mm}f_1)\,\right]&&  . \label{RIX4}
\end{eqnarray}
In the latter equation the standard notations of the ordinary
Boltzmann kinetics are used and $a(f)$ is a positive and
increasing arbitrary function, which does not affect the
stationary form of $f$. The factor $G$ is the transition rate,
depending only on the nature of the two body particle
interaction.  The composition law
$\otimes\mbox{\raisebox{-2mm}{\hspace{-2.5mm}$\scriptstyle
\kappa$}}\hspace{1mm}$ defined through
\begin{eqnarray}
\ln_{_{\{{\scriptstyle
\kappa}\}}}(f\otimes\mbox{\raisebox{-2mm}{\hspace{-3.3mm}$\scriptstyle
\kappa$}} \hspace{2mm}h)= \ln_{_{\{{\scriptstyle
\kappa}\}}}f+\ln_{_{\{{\scriptstyle \kappa}\}}}h \ \ ,
\label{RIX5}
\end{eqnarray}
is a one-parameter generalization of the ordinary product having
the following properties
\\i) associative law:
$(f\otimes\mbox{\raisebox{-2mm}{\hspace{-3.3mm}$\scriptstyle
\kappa$}}
\hspace{2mm}h)\otimes\mbox{\raisebox{-2mm}{\hspace{-3.3mm}$\scriptstyle
\kappa$}}
\hspace{2mm}w=f\otimes\mbox{\raisebox{-2mm}{\hspace{-3.3mm}$\scriptstyle
\kappa$}} \hspace{2mm}
(h\otimes\mbox{\raisebox{-2mm}{\hspace{-3.3mm}$\scriptstyle
\kappa$}} \hspace{2mm}w)$,
\\ ii) neutral element:
$f\otimes\mbox{\raisebox{-2mm}{\hspace{-3.3mm}$\scriptstyle
\kappa$}}
\hspace{2mm}1=1\otimes\mbox{\raisebox{-2mm}{\hspace{-3.3mm}$\scriptstyle
\kappa$}} \hspace{2mm}f=f$,
\\ iii) inverse element:
$f\otimes\mbox{\raisebox{-2mm}{\hspace{-3.3mm}$\scriptstyle
\kappa$}} \hspace{2mm}(1/f)=
(1/f)\otimes\mbox{\raisebox{-2mm}{\hspace{-3.3mm}$\scriptstyle
\kappa$}} \hspace{2mm}f=1$,
\\ iv) commutative law: $f
\otimes\mbox{\raisebox{-2mm}{\hspace{-3.3mm}$\scriptstyle
\kappa$}}
\hspace{2mm}h=h\otimes\mbox{\raisebox{-2mm}{\hspace{-3.3mm}$\scriptstyle
\kappa$}} \hspace{2mm}f$. \\ Furthermore it results
$f\otimes\mbox{\raisebox{-2mm}{\hspace{-3.3mm}$\scriptstyle
\kappa$}}
\hspace{2mm}0=0\otimes\mbox{\raisebox{-2mm}{\hspace{-3.3mm}$\scriptstyle
\kappa$}} \hspace{2mm}f= 0$ and the generalized division
$\oslash\mbox{\raisebox{-2mm}{\hspace{-2.2mm}$\scriptstyle
\kappa$}}\hspace{.5mm}$ can be defined through
$f\oslash\mbox{\raisebox{-2mm}{\hspace{-3.3mm}$\scriptstyle
\kappa$}}\hspace{2mm}h=f
\otimes\mbox{\raisebox{-2mm}{\hspace{-3.3mm}$\scriptstyle
\kappa$}} \hspace{1mm}(1/h)$.

{\it H-theorem:} The four-vector entropy
$S^{\nu}=(S^{0},\mbox{\boldmath $S$})$ it was defined as
\begin{equation}
S^{\nu}= - \int \frac{d^3p}{p^{0}}\,p^{\nu}\, f
\,\ln_{_{\{{\scriptstyle \kappa}\}}}\!f \ \ . \label{RIX6}
\end{equation}
The temporal component $S^{0}$ of this four-vector coincides with
the $\kappa$-entropy defined previously through Eq. (\ref{RVIII1})
while the spatial component $\mbox{\boldmath $S$}$ is the entropy
flow. Starting from the evolution equation (\ref{RIX4}) it has
been demonstrated the H-theorem, which represents the second law
of thermodynamics and states that the entropy production is never
negative, and, in equilibrium conditions, there is no entropy
production. Namely, the following relation was obtained
\begin{equation}
\partial_{\nu}S^{\nu}\geq 0 \ \ , \label{RIX7}
\end{equation}
which is the local formulation of the relativistic H-theorem and
represents the second law of the thermodynamics \cite{PRE02}.

{\it Entropy transformation:} After recalling the identity
$d^3p/p^0=d^4p \,\, 2\,
\theta(p^0)\,\delta(p^{\mu}p_{\mu}-m^2c^2)$ the definition
(\ref{RIX6}) assumes the form
\begin{equation}
S^{\nu}= - \int
d^4p\,\,2\,\theta(p^{0})\,\delta(p^{\mu}p_{\mu}-m^2c^2)\,\,p^{\nu}\,
f \,\ln_{_{\{{\scriptstyle \kappa}\}}}\!f \ \ . \label{RIX8}
\end{equation}
In this expression $d^4p$ is a scalar because the Jacobian of the
Lorentz transformation is equal to unit. Then since $p^{\nu}$
transforms as a four-vector, we can conclude that $S^{\nu}$
transforms as a four-vector like the particle four-flow
\begin{equation}
N^{\nu}= \int
d^4p\,\,2\,\theta(p^{0})\,\delta(p^{\mu}p_{\mu}-m^2c^2)\,\,p^{\nu}\,
f \ \ , \label{RIX9}
\end{equation}
for which the conservation law of the total particle number
imposes
\begin{equation}
\partial_{\nu}N^{\nu}= 0 \ \ . \label{RIX10}
\end{equation}

{\it Calculation of the parameter $\kappa$:} We focus now
attention to the general discussion of sec. II. The results
obtained in sect. II permit us to consider the problem related to
the calculation of the parameter $\kappa$ from a very different
and more physically sound point of view with respect the one of
ref. \cite{PRE02}. Clearly, the velocity $v_*$ introduced through
$|\kappa|=v_*/c$ does not emerge within the one-particle
relativistic theory.  In general $v_*$ could depend on $c$ namely
$v_*=v_*(c)$ with the condition that $v_*(\infty)<\infty$. The
fact that the condition $|\kappa|<1$ and then $v_*<c$, follows
from the maximum entropy principle, supports the supposition that
$v_*$ could emerge in the framework of the many body theory.
Taking in mind that in statistical mechanics the only emerging
velocity is the thermal velocity, in the following we identify
$v_*$ with the thermal velocity of the many body system.

The particle thermal energy, being the mean particle kinetic
energy $\langle W\rangle$, is related to the thermal velocity
$v_*=\kappa c$ according to the relativistic relationship
\begin{equation}
\langle W \rangle\,=
mc^2\left(\frac{1}{\sqrt{1-\kappa^2}}-1\right) \ \ . \label{RIX11}
\end{equation}
Clearly, in a statistical theory, the mean particle kinetic energy
$\langle W \rangle$ is considered a known quantity. Then the
expression of $\kappa$ in terms of $\langle W \rangle$ follows
immediately
\begin{equation}
\kappa^2=1-\bigg(1+\frac{\langle W \rangle}{mc^2}\bigg)^{-2} \ \ ,
\label{RIX12}
\end{equation}
so that the theory does not contain free parameters and in the
classical limit $c\rightarrow \infty$, results $\kappa=0$. From
(\ref{RIX12}) follows that $\lim_{c\rightarrow \infty}\, \kappa
c\!=\!\sqrt{2<W>/m}<\infty$ as required from the theory developed
in sect. II. Note that in the case we know experimentally the
distribution function, the value of $\kappa$ and consequently, by
using (\ref{RIX12}), the value of the ratio $\langle W
\rangle/mc^2$ can be obtained from the slope of the distribution
tails in a log-log plot.  The normalization condition for the
distribution function $f$ and the definition of the particle mean
kinetic energy, starting from $f$, permit us to determine
univocally $T$ and $\mu$ . In this way one can link $\langle
W\rangle$ and then $\kappa$ to the temperature of the system.

Bearing in mind the meaning of the parameter $\kappa$ we rewrite
the distribution function (\ref{RVIII7}) in the form
\begin{equation}
f=\alpha \exp_{_{\{{\scriptstyle \kappa}\}}}\!\!\left(-\,
\frac{W-\mu}{k_{_{B}}T'}\right) \ \ , \label{RIX13}
\end{equation}
being $W$ the relativistic kinetic energy while the temperature
$T'$ is given by
\begin{equation}
T'=\sqrt{1-\frac{v_*^{\,2}}{c^{\,2}}}\,\, T \ \ . \label{RIX14}
\end{equation}
We remark that the temperature $T'$ is contracted with respect to
the temperature T by a factor which, curiously, has the same form
of the one entering in the definition of the relativistic
temperature suggested by Einstein \cite{Einstein} and Planck
\cite{Planck}.

\sect{Applications}

In the last four years the present theory has been used
successfully to study some systems which manifestly cannot be
treated within the ordinary statistical mechanics. Clearly, the
most natural applications regard relativistic many body systems.

The first application, considered in ref. \cite{PRE02}, concerns
the cosmic rays. Since long time it is known that the cosmic rays
spectrum, which extends over $13$ decades in energy, from a few
hundred of MeV ($10^8$ eV) to a few hundred of EeV ($10^{20}$ eV)
and spans $33$ decades in particle flux, from $10^{4}$ to
$10^{-29}$ $[m^2\, sr\, s\, GeV]^{-1}$, is not exponential and,
then, it violates the Boltzmann equilibrium statistical
distribution $\propto \exp (-E/k_{_{B}}T)$
\cite{Vasyliunas,Biermann,Swordy}. On the other hand, it is known
that the particles composing the cosmic rays are essentially the
normal nuclei as in the standard cosmic abundance of matter. Then,
the cosmic rays can be viewed as an equivalent statistical system
of identical relativistic particles with masses near the mass of
the proton ($938$ MeV). These characteristics (relativistic
particles with a very large extension both for their flux and
energy) yield the cosmic rays spectrum an ideal physical system
for a preliminary test of the correctness and predictability of
any relativistic theory. In ref. \cite{PRE02},  a high quality
agreement between the predictions of the present theory and the
observed data, in the whole cosmic ray spectrum, has been found.
This agreement over so many decades is quite remarkable.

Other physical applications of the theory regard the formation of
quark-gluon plasma \cite{quarkgluon} and kinetic models
describing a gas of interacting atoms and photons \cite{Rossani}.

On the other hand, the theory has been applied successfully also
to the study of natural or artificial systems exhibiting a
limiting velocity in the propagation of the information (like the
one imposed by light speed in physics) where a mechanism
analogous to the relativistic one can emerge, deforming the
distribution function. In ref. \cite{Fracture} the problem of the
fracture propagation has been considered within the present
theory. Others applications have been considered in games theory
\cite{Topsoe}, in economy for study the income distribution
\cite{Dragul} and in construct financial models \cite{econlett},
etc.

We come back to the high energy physics being the more proper
field of application of the theory. In the following, as further
application, we consider the open problem of the black hole
physics regarding the intriguing question of the
Bekenstein-Hawking area law \cite{Bekenstein,Hawking}. This law
asserts that the entropy of a black hole scales as the area $A$
of the event horizon namely $S\propto A$. It is well known that
the ordinary Bolzmann statistical mechanics and thermodynamics
fails to justify this law and in contrary predicts that the
entropy scales as the volume $V$ of spatial region delimited by
the event horizon, namely $S\propto V$.

Indeed, if we indicate with $N$ the degrees of freedom or
microscopic components in the volume $V$ and with $M$ the number
of states accessible by each component, within the Boltzmann
statistics the number of microstates $\Omega$ of the whole system
is given by $\Omega \propto M^N$ with $N>>1$. After introducing
the notations
\begin{eqnarray}
n(0)=\exp N \ \ , \ \ m(0)=\ln M \ \ , \label{RX1}
\end{eqnarray}
the number of microstates can be written also in the form
\begin{eqnarray}
\Omega\propto n(0)^{m(0)} \ \ . \label{RX2}
\end{eqnarray}
Customary one assumes a uniform distribution of the components in
the volume $V$ obtaining $N\propto V$. Thus the entropy
\begin{eqnarray}
S=\ln \Omega \ \ , \label{RX3}
\end{eqnarray}
of the black hole results to be given by
\begin{eqnarray}
S \propto V \ \ . \label{RX4}
\end{eqnarray}
This result of the Boltzmann statistics, already known in the
literature, is manifestly in contradiction with the
Bekenstein-Hawking area law.

Very recently, in ref. \cite{Botta}, has been suggested to treat
this problem within non-standard statistics. Even if the authors
limit their discussion to the nonextensive \cite{Gell-Mann}
aspects of the problem the proposed approach has a general
validity and can be adopted also in the present case.

Clearly the black hole is a relativistic system and appears
natural to write its entropy in the microcanonical picture as
\begin{eqnarray}
S=\ln_{_{\{{\scriptstyle \kappa}\}}} \Omega \ \ . \label{RX5}
\end{eqnarray}
Furthermore the present formalism generalizes Eqs. (\ref{RX1}) and
(\ref{RX2}) as follows
\begin{eqnarray}
n(\kappa)=\exp_{_{\{{\scriptstyle \kappa}\}}}\! N \ \ , \ \
m(\kappa)=\ln_{_{\{{\scriptstyle \kappa}\}}}\! M \ \ , \label{RX6}
\end{eqnarray}
\begin{eqnarray}
\Omega\propto n(\kappa)^{m(\kappa)} \ \ . \label{RX7}
\end{eqnarray}
After recalling the property (\ref{RA15}) the number of
microstates becomes
\begin{eqnarray}
\Omega\propto\exp_{_{\{{\scriptstyle
\kappa/m(\kappa)}\}}}\!\!\big(m(\kappa)N\big) \ \ , \label{RX8}
\end{eqnarray}
while the black hole entropy can be written as
\begin{eqnarray}
S\propto\ln_{_{\{{\scriptstyle
\kappa}\}}}\!\!\left[\,\exp_{_{\{{\scriptstyle
\kappa/m(\kappa)}\}}}\!\!\big(m(\kappa)N\big)\right] \ \ .
\label{RX9}
\end{eqnarray}
In the thermodynamic limit $N>>1$, then Eqs. (\ref{RX8}) and
(\ref{RX9}) predict that the number of microstates and the
entropy scales as power laws
\begin{eqnarray}
&&\Omega\propto N^{m(\kappa)/\kappa} \ \ , \label{RX10} \\
&&S\propto N^{m(\kappa)}  \ \ . \label{RX11}
\end{eqnarray}
We remark that the power law behavior  of the various physical
quantities is a feature of the present theory in contrast with the
exponential and logarithmic behaviors in classical statistical
mechanics. Finally, we assume,  as customary, a uniform
distribution of the components in the volume $V$, namely $N
\propto V$ and obtain the following expression of entropy
\begin{eqnarray}
S\propto V^{m(\kappa)} \ \ . \label{RX12}
\end{eqnarray}
After posing
\begin{eqnarray}
m(\kappa)=2/3 \ \ , \label{RX13}
\end{eqnarray}
the Bekenstein-Hawking law, namely  $S\propto A$, follows
immediately and thus we can conclude that the present statistical
theory is consistent with the area law of black hole entropy. This
result is in agreement with ref. \cite{Botta} where the authors
propose a relationship analogous to the one given by Eq.
(\ref{RX13}).

\sect{Epimythion}

We have shown that within the standard framework of the special
relativity, the Lorentz transformations lead to a proper
one-parameter generalization of the entropy just as for other
physically meaningful quantities (e.g. momentum, energy, etc).

The new entropy generates a coherent and selfconsistent
relativistic statistical theory (both statistical mechanics as
well as kinetic theory) preserving the main features (maximum
entropy principle, thermodynamic stability, continuity condition,
H-theorem, etc) of the ordinary statistical theory  which
riemerges in the classical limit.

The distribution function predicted by the present theory is a
one-parameter, continuous deformation of the classical
Maxwell-Boltzmann distribution exhibiting a power law tail in
accordance with the experimental evidence in several relativistic
many body systems.

\sect{Appendix}

We start by considering the $\kappa$-sum for the dimensionless
relativistic momenta
\begin{eqnarray}
t_1\stackrel{\kappa}{\oplus}t_2=t_1\sqrt{1+\kappa^2t_2^2}
+t_2\sqrt{1+\kappa^2t_1^2} \ \ , \label{RA1}
\end{eqnarray}
and after posing $t_1=t+dt$ and  $t_2=-t$ we can introduce
$\kappa$-differential through
\begin{eqnarray}
d_{{\scriptscriptstyle\{}\kappa{\scriptscriptstyle\}}}t\!\!\!\!&&
=(t+dt)
\stackrel{\kappa}{\ominus}t \nonumber \\
&&= (t+dt) \stackrel{\kappa}{\oplus}(-t) \nonumber \\
&&=\frac{d t}{\sqrt{1+\kappa^2 t^2 }} \ \ . \label{RA2}
\end{eqnarray}
The $\kappa$-derivative
$d/d_{{\scriptscriptstyle\{}\kappa{\scriptscriptstyle\}}}t$
admits as eigenfunction the $\kappa$-exponential, namely
\begin{equation}
\frac{d\,}{d_{_{\{{\scriptstyle
\kappa}\}}}t}\,\exp_{_{\{{\scriptstyle
\kappa}\}}}(t)=\exp_{_{\{{\scriptstyle \kappa}\}}}(t) \ \ ,
\label{RA3}
\end{equation}
being
\begin{eqnarray}
&&\exp_{_{\{{\scriptstyle \kappa}\}}}(t)= \left(\sqrt{1+\kappa^2
t^{\,2}}+\kappa t\right)^{1/\kappa}\ \ , \label{RA4}
\\
&&\exp_{_{\{{\scriptstyle \kappa}\}}}\!\left(t
\right)=\exp\left(\frac{1}{\kappa}\,{\rm arcsinh}(\kappa t)\right)
\ \ . \label{RA5}
\end{eqnarray}
The $\kappa$-exponential can be viewed as a generalization of the
ordinary exponential
\begin{eqnarray}
&&\exp_{_{\{{\scriptstyle 0}\}}}(t)=\exp (t) \ \ , \label{RA6}
\\
&&\exp_{_{\{{\scriptstyle - \kappa}\}}}(t)=\exp_{_{\{{\scriptstyle
\kappa}\}}}(t) \ \ , \label{RA7}
\end{eqnarray}
and has the properties
\begin{eqnarray}
&&\exp_{_{\{{\scriptstyle \kappa}\}}}(t) \in C^{\infty}({\bf R})
\ \ , \label{RA8}\\
&&\frac{d}{d\,t}\, \exp_{_{\{{\scriptstyle \kappa}\}}}(t)>0
\ \ , \label{RA9}\\
&&\frac{d^2}{d\,t^2}\, \exp_{_{\{{\scriptstyle \kappa}\}}}(t)>0
\ \ , \label{RA10}\\
&&\exp_{_{\{{\scriptstyle \kappa}\}}}(-\infty)=0^+
\ \ , \label{RA11}\\
&&\exp_{_{\{{\scriptstyle \kappa}\}}}(0)=1
\ \ , \label{RA12}\\
&&\exp_{_{\{{\scriptstyle \kappa}\}}}(+\infty)=+\infty
\ \ , \label{RA13}\\
&&\exp_{_{\{{\scriptstyle \kappa}\}}}(t)\exp_{_{\{{\scriptstyle
\kappa}\}}}(-t)= 1 \ \ . \label{RA14}
\end{eqnarray}
Furthermore it results
\begin{eqnarray}
&&\!\!\!\!\!\!\!\left (\exp_{_{\{{\scriptstyle
\kappa}\}}}(t)\right )^{r} =\exp_{_{\{{\scriptstyle
\kappa/r}\}}}(r t) \ \ , \label{RA15}
\\
&&\!\!\!\!\!\!\!\exp_{_{\{{\scriptstyle
\kappa}\}}}(t_1)\exp_{_{\{{\scriptstyle
\kappa}\}}}(t_2)=\exp_{_{\{{\scriptstyle
\kappa}\}}}(t_1\stackrel{\kappa}{\oplus}t_2) \ \ , \label{RA16}
\\
&&\!\!\!\!\!\!\!\exp_{_{\{{\scriptstyle \kappa}\}}}(t_1) \!
\otimes\mbox{\raisebox{-2mm}{\hspace{-3.0mm}$\scriptstyle
\kappa$}} \hspace{2mm}\! \!\exp_{_{\{{\scriptstyle
\kappa}\}}}(t_2)=\exp_{_{\{{\scriptstyle \kappa}\}}}(t_1+t_2) \ \
, \label{RA17}
\end{eqnarray}
where the product
$\otimes\mbox{\raisebox{-2mm}{\hspace{-2.2mm}$\scriptstyle
\kappa$}} \hspace{2mm}$ is defined through Eq. (\ref{RIX5}). The
following asymptotic relations also hold
\begin{eqnarray}
\exp_{_{\{{\scriptstyle
\kappa}\}}}(t){\atop\stackrel{\textstyle\sim}{\scriptstyle
t\rightarrow
{0}}}\!\!\!\!\!\!\!\!&&1+t+\frac{t^2}{2}+(1-\kappa^2)\frac{t^3}{3!}
\ \ , \label{RA18}
\end{eqnarray}
\begin{eqnarray}
\exp_{_{\{{\scriptstyle
\kappa}\}}}(t){\atop\stackrel{\textstyle\sim}{\scriptstyle
t\rightarrow {\pm\infty}}}\!\!\!\!\!\!\!\!&&|2\kappa t
|^{\pm1/|\kappa|} \ \ . \label{RA19}
\end{eqnarray}
Finally, after tedious but straightforward calculations the
following formula can be obtained
\begin{equation}
\int_{\, 0}^{\infty}\!\!t^{r\!-\!1}\!\exp_{_{\{{\scriptstyle
\kappa}\}}}\!(\!-t)\,dt =\frac{|2\kappa|^{-r}}{1\!+\!r
|\kappa|}\,\,\,
\frac{\Gamma\!\!\left(\frac{1}{|2\kappa|}\!-\!\frac{r}{2}
\right)}{\Gamma\!\!\left(\frac{1}{|2\kappa|}\!+\!\frac{r}{2}
\right)}\,\,\Gamma\!\!\left(r\right) \ \ , \label{RA20}
\end{equation}
holding for $r|\kappa|<1$. Note that this integral is given also
in \cite{PRE02} by Eq. (5.31) which unfortunately contains a
typing error (the first minus in the denominator of the right hand
side must be replaced by a plus so that after some simple algebra
Eq. (5.31) of \cite{PRE02} reproduces the present result).

The $\kappa$-logarithm $\ln_{_{\{{\scriptstyle \kappa}\}}}(t)$,
defined as the inverse function of the of $\kappa$-exponential,
namely  $\ln_{_{\{{\scriptstyle
\kappa}\}}}(\exp_{_{\{{\scriptstyle
\kappa}\}}}t)=\exp_{_{\{{\scriptstyle
\kappa}\}}}(\ln_{_{\{{\scriptstyle \kappa}\}}}t)=t$, is given by
\begin{eqnarray} &&\ln_{_{\{{\scriptstyle \kappa}\}}}(t)=
\frac{t^{\kappa}-t^{-\kappa}}{2\kappa} \ \ , \label{RA21}
\\
&&\ln_{_{\{{\scriptstyle \kappa}\}}}\!\left(t
\right)=\frac{1}{\kappa}\sinh (\kappa \ln t) \ \ , \label{RA22}
\end{eqnarray}
and can be viewed as a generalization of the ordinary logarithm
\begin{eqnarray}
&&\ln_{_{\{{\scriptstyle 0}\}}}(t)=\ln (t)\ \ , \label{RA23}
\\
&&\ln_{_{\{{\scriptstyle - \kappa}\}}}(t)=\ln_{_{\{{\scriptstyle
\kappa}\}}}(t) \ \ . \label{RA24}
\end{eqnarray}
The $\kappa$-logarithm, just as the ordinary logarithm, has the
properties
\begin{eqnarray}
&&\ln_{_{\{{\scriptstyle \kappa}\}}}(t) \in C^{\infty}({\bf R}^+)
\ \ , \label{RA25}\\
&&\frac{d}{d\,t}\, \ln_{_{\{{\scriptstyle \kappa}\}}}(t)>0
\ \ , \label{RA26}\\
&&\frac{d^2}{d\,t^2}\, \ln_{_{\{{\scriptstyle \kappa}\}}}(t)<0
\ \ , \label{RA27}\\
&&\ln_{_{\{{\scriptstyle \kappa}\}}}(0^+)=-\infty
\ \ , \label{RA28}\\
&&\ln_{_{\{{\scriptstyle \kappa}\}}}(1)=0
\ \ , \label{RA29}\\
&&\ln_{_{\{{\scriptstyle \kappa}\}}}(+\infty)=+\infty
\ \ , \label{RA30}\\
&&\ln_{_{\{{\scriptstyle \kappa}\}}}(1/t)=-\ln_{_{\{{\scriptstyle
\kappa}\}}}(t) \ \ . \label{RA31}
\end{eqnarray}
Moreover it has also the properties
\begin{eqnarray}
&&\ln_{_{\{{\scriptstyle \kappa}\}}}(t^{r}) =r
\ln_{_{\{{\scriptstyle r \kappa}\}}}(t) \ \ , \label{RA32}
\\
&&\ln_{_{\{{\scriptstyle \kappa}\}}}(t_1 \,t_2)
=\ln_{_{\{{\scriptstyle \kappa}\}}}(t_1)
\oplus\!\!\!\!\!^{^{\scriptstyle
\kappa}}\,\,\ln_{_{\{{\scriptstyle \kappa}\}}}(t_2)\ \ ,
\label{RA33} \\
&&\ln_{_{\{{\scriptstyle \kappa}\}}}(t_1\!
\otimes\mbox{\raisebox{-2mm}{\hspace{-3.0mm}$\scriptstyle
\kappa$}} \hspace{2mm}\! t_2) =\ln_{_{\{{\scriptstyle
\kappa}\}}}(t_1) + \ln_{_{\{{\scriptstyle \kappa}\}}}(t_2)\ \ ,
\label{RA34}
\end{eqnarray}
while the following asymptotic formulas hold
\begin{eqnarray}
\ln_{_{\{{\scriptstyle \kappa}\}}}(1+t)
{\atop\stackrel{\textstyle\sim}{\scriptstyle t\rightarrow {0}}}
t-\frac{t^2}{2}+\left(1+\frac{\kappa^2}{2}\right)\frac{t^3}{3} \
\ , \label{RA35}
\end{eqnarray}
\begin{eqnarray}
&&\ln_{_{\{{\scriptstyle
\kappa}\}}}(t){\atop\stackrel{\textstyle\sim}{\scriptstyle
t\rightarrow {0^+}}}-\frac{1}{2|\kappa|}\,\,t^{-|\kappa|} \ \ ,
\label{RA36}
\\ &&\ln_{_{\{{\scriptstyle
\kappa}\}}}(t){\atop\stackrel{\textstyle\sim}{\scriptstyle
t\rightarrow {+\infty}}}\,\frac{1}{2|\kappa|}\,\,t^{\,|\kappa|} \
\ . \label{RA37}
\end{eqnarray}
Furthermore holds the very useful formula
\begin{eqnarray}
\frac{d^2}{d\,t^2}\,t\, \ln_{_{\{{\scriptstyle \kappa}\}}}(t)> 0
\ \ , \label{RA38}
\end{eqnarray}
and finally one obtains \cite{PRE02}
\begin{equation}
\int_{\, 0}^{1}dt \left(\ln_{_{\{{\scriptstyle
\kappa}\}}}\!\frac{1}{t}\right)^{r-1}
\!\!\!\!\!=\!\frac{|2\kappa|^{1-r}}{1+(r-1)|\kappa|}\,\,\,
\frac{\Gamma\!\left(\!\frac{1}{|2\kappa|}\!-\!\frac{r-1}{2}
\right)}{\Gamma\!\left(\!\frac{1}{|2\kappa|}\!+\!\frac{r-1}{2}
\right)} \,\,\Gamma\!\left(r\right) \ \ . \label{RA39}
\end{equation}

It results evident that the above used procedure, to introduce the
$\kappa$-exponential and $\kappa$-logarithm, starting from the
additivity law of the momenta, can be used to obtain new functions
if we start from other generalized additivity laws. For instance,
starting from the additivity law of the velocities we have
$d_{{\scriptscriptstyle\{}\kappa{\scriptscriptstyle\}}}t=(1-\kappa^2t^2)^{-1}dt$,
while instead of $\kappa$-exponential we obtain the function
\begin{eqnarray}
\varphi(t)\!\!\!&&= \left(\frac{1+\kappa t}{1-\kappa
t}\right)^{1/2\kappa} \ \ , \label{RA40}
\\
&&=\,\exp\left(\frac{1}{\kappa}{\rm arctanh}(\kappa t)\right) \ \
, \label{RA41}
\end{eqnarray}
whose inverse function assumes the form
\begin{eqnarray}
\varphi^{-1}(t)\!\!\!&&=
\frac{1}{\kappa}\,\,\frac{t^{\kappa}-t^{-\kappa}}{t^{\kappa}+t^{-\kappa}}
\ \ , \label{RA42} \\ &&=\frac{1}{\kappa}\tanh (\kappa \ln t) \ \
. \label{RA43}
\end{eqnarray}

The function $\varphi(t)\in C^{\infty}(I)$ with
$I=]-|\kappa|^{-1}\, , \,+|\kappa|^{-1}[$ is connected to
$\exp_{_{\{{\scriptstyle \kappa}\}}}(t)$ through
\begin{eqnarray}
\varphi(t)=\exp_{_{\{{\scriptstyle
\kappa}\}}}\!\left(\frac{t}{\sqrt{1-\kappa^2 t^2}}\right) \ \ .
\label{RA44}
\end{eqnarray}
At this point one could be tempted to replace
$\exp_{_{\{{\scriptstyle \kappa}\}}}(t)$ with the function
$\varphi(t)$ as the right one for generalize the ordinary
exponential.

There are several reasons to consider $\exp_{_{\{{\scriptstyle
\kappa}\}}}(t)$ as the more proper generalization of the ordinary
exponential. Beside the motivations emerging within the
one-particle relativistic dynamics, there is the one related to
the maximum entropy principle. This principle, the cornerstone of
statistical physics, selects without any ambiguity (see sect. VII)
just the $\kappa$-exponential as the more natural generalization
of the ordinary exponential. More precisely the function
$\exp_{_{\{{\scriptstyle \kappa}\}}}(t)$ is the only one existing
which  emerges simultaneously both in the one-particle
relativistic dynamics as well as in the many-body physics through
the maximum entropy principle.

\end{document}